\documentclass[aps,pra,twocolumn]{revtex4-2}
\usepackage{amsfonts}
\usepackage{amsmath}
\usepackage{mathrsfs}
\usepackage{amssymb}
\usepackage{graphicx}
\usepackage{bm}
\usepackage{color}
\usepackage{CJK}
\usepackage{dcolumn}
\usepackage{verbatim}
\usepackage{float}
\usepackage{booktabs}   
\usepackage{array}    
\usepackage{multirow}  
\usepackage{threeparttable}  
\usepackage[colorlinks=true,linkcolor=blue,anchorcolor=blue,citecolor=blue,urlcolor=blue]{hyperref}

\begin{document}
\preprint{APS/123-QED}

\title{Simulation of chemical reaction dynamics based on quantum computing}

\author{Qiankun Gong,$^1$ Qingmin Man,$^1$ Ye Li,$^1$ Menghan Dou,$^1$ Qingchun Wang,$^{2,*}$ \\
Yu-Chun Wu,$^{2,3,4,5}$ Guo-Ping Guo$^{1,2,3,4,5,}$} 
\email[]{qingchun720@ustc.edu.cn;\\gpguo@ustc.edu.cn}

\affiliation{$^1$Origin Quantum Computing Company Limited, Hefei 230026, China \\
$^2$Institute of Artificial Intelligence, Hefei Comprehensive National Science Center, Hefei, Anhui 230088, China \\
$^3$CAS Key Laboratory of Quantum Information, School of Physics, University of Science and \\Technology of China, Hefei, Anhui, 230026, China \\
$^4$CAS Center For Excellence in Quantum Information and Quantum Physics, University of Science and \\Technology of China, Hefei, Anhui, 230026, China \\ 
$^5$Hefei National Laboratory, University of Science and Technology of China, Hefei 230088, China
}

\begin{abstract}
The molecular energies of chemical systems have been successfully calculated on quantum computers, however, more attention has been paid to the dynamic process of chemical reactions in practical application, especially in catalyst design, material synthesis. Due to the limited the capabilities of the noisy intermediate scale quantum (NISQ) devices, directly simulating the reaction dynamics and determining reaction pathway still remain a challenge. Here we develop the ab initio molecular dynamics based on quantum computing to simulate reaction dynamics by extending correlated sampling approach. And, we use this approach to calculate Hessian matrix and evaluate computation resources. We test the performance of our approach by simulating hydrogen exchange reaction and bimolecular nucleophilic substitution S$\rm_N$2 reaction. Our results suggest that it is reliable to characterize the molecular structure, property, and reactivity, which is another important expansion of the application of quantum computing.

\end{abstract}

\maketitle

\section{Introduction}
Reactive collision, which involves the chemical transformation, is regarded as the central phenomenon in physical chemistry\cite{guo2012quantum,zhang2016recent}. The reaction dynamics usually occurs in the femtosecond scale, involving the breakage of old bonds and the formation of new bonds. Therefore, the quantum mechanics is necessary to comprehensively describe and character the reactive event.  The H + H$_2$ reaction dynamics had been simulated by full quantum mechanical calculation about forty years ago\cite{kuppermann1975quantum,karplus2014development}. And many chemical reactions have been studied using quantum mechanics methods with the increase of computer power. However, the reaction dynamics is still limited to small molecular systems due to the exponential increment of quantum mechanics in computational complexity\cite{guo2012quantum,fu2018ab}. Quantum computer, a new computing paradigm utilizing the superposition and entanglement of qubits, has powerful processing speed, and is expected to bring new solutions to the electronic structure problems\cite{mcardle2020quantum,cao2019quantum,barkoutsos2021quantum,ma2020quantum}. Therefore, some quantum algorithms have been developed to simulate the molecular properties, such geometry optimization, response properties, vibrational structure and so on\cite{kassal2009quantum,delgado2021variational,o2019calculating,mitarai2020theory,ollitrault2020hardware,sawaya2021near,lotstedt2021calculation,cai2020quantum,christensen2019operators,oftelie2022computing,di2022quantum}.

According to quantum computer framework, some dynamics approaches based on quantum algorithms have been proposed to investigate the chemical process\cite{kassal2008polynomial,ollitrault2020nonadiabatic,ollitrault2021molecular,macdonell2021analog,langkabel2022quantum,fedorov2021ab,sokolov2021microcanonical}. For exanmple,  Kassal, Ollitrault and MacDonell simulated the real-time evolution of nonadiabatic chemical processes on real quantum devices or digital simulators respectively\cite{kassal2008polynomial,ollitrault2021molecular,macdonell2021analog}. However, the circuit depth and coherence time required by their approaches, as well as the subsequent error correction on quantum hardware, aren’t achievable in present noisy intermediate-scale quantum (NISQ) era. Consequently, Sokolov and Fedorov proposed the correlated sampling approach by combining ab initio molecular dynamics (AIMD) and variational quantum eigensolver (VQE) algorithm, and tested the framework by simulating the simplest vibration of H$_2$ molecule\cite{fedorov2021ab}. In AIMD simulation, Born–Oppenheimer (or adiabatic) approximation is adopted, which assumes that the wave function of the whole system can be regarded as a simple product of the nuclear and electronic wave functions in molecular systems, and the crossing of potential energy surfaces (PES) is not considered, thus one can deal with the motion of electrons and nuclei independently\cite{karplus2014development,karplus1965exchange} Compared to full quantum molecular dynamics and classical molecular dynamics, the AIMD achieve a good balance between speed and accuracy. 

In this study, we first extend the correlated sampling approach to general cases, then simulate the H$^+$ + H$_2$ → H$_2$  + H$^+$  reaction dynamics by AIMD simulation based on VQE algorithm. In the extended approach, the measurement expectation can be reused if the Pauli string operator of force is the same as the ground-state energy, otherwise one need measure the Pauli string operator based on the same ground-state wave function, which minimizes the quantum calculation resources in AIMD simulation. To test the approach in more complex system, the bimolecular nucleophilic substitution S$\rm_N$2 (Cl$^-$ + CH$_3$Cl → ClCH$_3$ + Cl$^-$) reaction is also simulated. Moreover, we investigate the Hessian matrix based on the correlated sampling approach. The results indicate that the computation complexity of Hessian matrix can be minimized when ignoring wave function derivatives with nuclear coordinates. Final, we locate the transition structure of H$^+$ + H$_2$ reaction and S$\rm_N$2 reaction based on the Hessian matrix.

This paper is organized as follows. In the section \ref{sec2} “Methods” part, we describe the calculation method of the ground-state energy, force, and Hessian matrix. In the section \ref{sec3} “Result and discussion” part
, we display the simulated results for the hydrogen exchange reaction and S$\rm_N$2 reaction.

\section{Methods}
\label{sec2}
 The VQE algorithm is a hybrid quantum–classical approach\cite{peruzzo2014variational,o2016scalable,kandala2017hardware,grimsley2019adaptive,google2020hartree}, in which the preparation and measurement of a parametrized ansatz are carried out on the quantum computer, and these parameters are iteratively optimized on a classical digital computer according to the Rayleigh–Ritz variational principle
\begin{equation}
	\label{eq1}
	E \leq\left\langle\Psi(\boldsymbol{\theta})\left|\hat{\mathrm{H}}_{e l}(\boldsymbol{R})\right| \Psi(\boldsymbol{\theta})\right\rangle,
\end{equation}
assuming the trial state is normalized, $\boldsymbol{\theta}$ is the variational parameters, $E$ is the ground-state energy, $\boldsymbol{R}$ is the coordinates of molecule. The second-quantized electronic Hamiltonian is
\begin{equation}
	\label{eq2}
	\hat{H}_{\mathrm{e} l}=\sum_{\mathrm{pq}} h_{p q} a_p^{\dagger} a_q+\frac{1}{2} \sum_{\mathrm{pqrs}} h_{p q r s} a_p^{\dagger} a_q^{\dagger} a_r a_s,
\end{equation}	
 where $h_{p q}$ and $h_{p q r s}$ is the singles- and doubles- electron integrals computed by classical computer respectively. $a^{\dagger}$ and $a$ are the creation and annihilation operators for corresponding spin orbitals.

 To simulate the second-quantized Hamiltonian, the creation and annihilation operators need to be encoded into qubits. The Jordan-Wigner (JW) transformation, parity transformation and Bravyi-Kitaev (BK) transformation\cite{bravyi2002fermionic} are the most popular encoding approaches now. Among them, JW transformation is a basic and widely used encoding, and the orbital occupation status corresponds to the qubit states. In this study, the JW transformation is used to produce the Pauli string operator formation of electronic Hamiltonian
 \begin{equation}
 	\label{eq3}
    \hat{H}_{\mathrm{el}}=\sum_{\alpha} h_\alpha\left(\prod_k^N \hat{p}_k\right)_\alpha,
\end{equation}
 $h_\alpha$ relies on the $h_{p q}$ and $h_{p q r s}$, $\hat{p}_k$ is the Pauli operator $\{X, Y, Z, I\}$, $N$ is the number of qubit. Then the ground-state energy after optimizer
\begin{equation}
	\label{eq4}
	\mathrm{E}= \sum_{\alpha}h_{\boldsymbol{\alpha}}\left\langle\Psi(\boldsymbol{\theta})\left|\left(\prod_k^N \hat{p}_k\right)_{\boldsymbol{\alpha}}\right| \Psi(\boldsymbol{\theta})\right\rangle=\sum_{\alpha} h_{\boldsymbol{\alpha}} P_{\boldsymbol{\alpha}},
\end{equation}
$P_{\boldsymbol{\alpha}}$ is the measurement expectation of Pauli string operator.

 The ansatz method is also a significant concept in VQE algorithm. A good ansatz could decrease circuit depth and improve calculation precision. The common ansatz methods include unitary coupled cluster (UCC) ansatz\cite{yaris1964linked,yaris1965cluster,taube2006new,peruzzo2014variational}, symmetry-preserved (SP) ansatz\cite{gard2020efficient} and hardware-efficient ansatz\cite{kandala2017hardware}. Benefit from the success of classical coupled cluster method and its variants, the UCC ansatz is the well-known ansatz in quantum computing field. The advantages of symmetry-preserved ansatz is using 2-qubit block as basis element and suitable to particular molecular system. The hardware-efficient ansatz includes many layers and is suitable to run on NISQ hardware. Here we make use of the unitary coupled cluster with singles and doubles (UCCSD) ansatz method
 \begin{equation}
 	\label{eq5}
 	|\Psi(\boldsymbol{\theta})\rangle=e^{\hat{T}(\theta)-\hat{T}(\theta)^{\dagger}}|\Psi\rangle_{H F},
\end{equation}
$\hat{T}(\theta)^{\dagger}$ is the excitation operator and the initial state is Hartree-Fock state.

The ab initio molecular dynamics simulations (AIMD) can be performed to directly simulate the reaction dynamics. In AIMD simulations, the Born-Oppenheimer approximation is used to separate the motions of the nucleus and electron. To reduce error for integrating the motion equations, the fairly small time steps is necessary and the total energy conservation is often applied to evaluate the simulation quality. Since the velocity-Verlet algorithm can keep the total energy conservation over long time simulation\cite{verlet1967computer,swope1982computer}, the positions and velocities of nuclei are updated at per step by 
\begin{equation}
	\label{eq6}
r(t+\Delta t)=r(t)+v(t) \Delta t+\frac{1}{2 m} F(t) \Delta t^2,
\end{equation}
\begin{equation}
	\label{eq7}
	v(t+\Delta t)=v(t)+\frac{1}{2 m}(F(t)+F(t+\Delta t)) \Delta t,
\end{equation}
$r$, $v$, $t$, $\Delta t$, $m$, $F$ are coordinate, velocity, time, time step, mass and force respectively. The initial velocities and coordinates can be obtained according to actual conditions. Therefore, the calculation of forces is the key of AIMD simulations.

In the Hellmann-Feynman theorem, the force of atomic nucleus along with the $j$ direction equals to
\begin{equation}
	\label{eq8}
F_j=-\left\langle\Psi(\boldsymbol{\theta})\left|\frac{\partial \hat{H}_{e l}(\boldsymbol{R})}{\partial \boldsymbol{R}_j}\right| \Psi(\boldsymbol{\theta})\right\rangle.
\end{equation}
Here we employ the central finite difference approach to approximate the force
\begin{equation}
	\label{eq9}
	F_j=-\frac{\left\langle\Psi(\boldsymbol{\theta}) \left| \Delta H \right|\Psi(\boldsymbol{\theta})\right\rangle}{2 \times \Delta d},
\end{equation}
where
\begin{equation}
	\label{eq10}
	\Delta H=\sum_{\gamma} h_\gamma\left(\prod_k^N \hat{p}_k\right)_\gamma-\sum_\beta h_\beta\left(\prod_k^N \hat{p}_k\right)_\beta.
\end{equation}

If $\left(\prod_k^N \hat{p}_k\right)_\gamma$ or $\left(\prod_k^N \hat{p}_k\right)_\beta$ is the same as the $\left(\prod_k^N \hat{p}_k\right)_\alpha$ , the measure probability $P_{\boldsymbol{\alpha}}$ can be reused. Otherwise, additional measurements are necessary based on the same ground-state wave function. Therefore, in our extended correlated sampling approach, the force can be obtained 
\begin{equation}
	\label{eq11}
     F_j=-\frac{\Delta E_j^{\text {reuse }}+\Delta E_j^{\text {extra }}}{2 \times \Delta d},
\end{equation}
where
\begin{equation}
	\label{eq12}	
    \Delta E_j^{\text {reuse }}=\sum_{\gamma, \alpha} h_\gamma P_\alpha-\sum_{\beta, \alpha} h_\beta P_\alpha,
\end{equation}
\begin{equation}
	\label{eq13}
	\Delta E_j^{\text {extra }}= \left\langle\Psi(\boldsymbol{\theta}) \left| \hat{O} \right| \Psi(\boldsymbol{\theta})\right\rangle,
\end{equation}
and
\begin{equation}
	\label{eq14}
	\hat{O}= \sum_{\gamma \notin \alpha} h_\gamma\left(\prod_k^N \hat{p}_k\right)_\gamma - \sum_{\beta \notin \alpha} h_\beta\left(\prod_k^N \hat{p}_k\right)_\beta\
\end{equation}

 Transition state (TS) is the maximum energy point along the reaction pathway that connects two minimum points. TS is a first-order saddle point on potential energy surface, there is one and only one imaginary frequency. The normal mode analysis is usually used to evaluate the vibration frequency. In normal mode analysis, the harmonic approximation is applied to characterize the potential energy surface,  coupled and collective motions of atoms can be separated into individually motion.
 
In present work, searching transition state is based on Newton-Raphson method
\begin{equation}
	\label{eq15}
    \boldsymbol{R}_{k+1}=\boldsymbol{R}_{k}-{H}_{k}^{-1} \nabla {E}_k,
\end{equation}
${H}_{k}^{-1}$, ${E}_k$, $\boldsymbol{R}_{k}$ are the inverse of the Hessian matrix, gradient, coordinates at $k$th iteration respectively. In general, the Hessian matrix must have only one negative eigenvalue at each iteration to ensure the TS optimization toward the desired direction. As a result, controlling step size and direction are necessary. Here we use the simplest way to control the step size, namely, scaling the step size by factor 0.5 when the only one negative eigenvalue is vanished. 

The Hessian matrix involves the second order energy derivative
\begin{equation}
	\label{eq16}
	S_{j,i}=\frac{\partial^2 {E}(\boldsymbol{R})}{\partial \boldsymbol{R}_{\mathrm{j}} \partial \boldsymbol{R}_{i}}.
\end{equation}
Using central finite difference approach
\begin{equation}
	\label{eq17}
    S_{j, i}=\frac{E_1+E_2-E_3-E_4}{4 \times \Delta d^2},
\end{equation}

where
\begin{equation}
    E_1=E\left(\boldsymbol{R}+\Delta d \boldsymbol{e}_{i}, \boldsymbol{R}+\Delta d \boldsymbol{e}_{j}\right),
\end{equation}
\begin{equation}
    E_2=E\left(\boldsymbol{R}-\Delta d \boldsymbol{e}_{i}, \boldsymbol{R}-\Delta d \boldsymbol{e}_{j}\right),
\end{equation}
\begin{equation}
    E_3=E\left(\boldsymbol{R}-\Delta d \boldsymbol{e}_{i}, \boldsymbol{R}+\Delta d \boldsymbol{e}_{j}\right),
\end{equation}
\begin{equation}
    E_4=E\left(\boldsymbol{R}+\Delta d \boldsymbol{e}_{i}, \boldsymbol{R}-\Delta d \boldsymbol{e}_{j}\right).
 \end{equation}
 When $i$ is equal to $j$
\begin{equation}
	\label{eq22}
	{S}_{{i}, {i}}=\frac{{E}\left(\boldsymbol{R}+2 \Delta {d} \boldsymbol{e}_{i}\right) - E\left(\boldsymbol{R}-2 \Delta {d} \boldsymbol{e}_{i}\right)}{4 \times \Delta {d}^2}.
\end{equation}

Compared to ground-state energy calculation, the Hessian matrix need optimizer $18N^2$ ground-state ware functions for $N$ atoms system, which is unrealistic for quantum devices. Inspired by the calculation of forces, the second order energy derivative can be approximated
\begin{equation}
	\label{eq23}
    {S}_{{j}, {i}}=\left\langle\Psi(\boldsymbol{R})\left|\frac{\partial^2 \widehat{{H}}(\boldsymbol{R})}{\partial \boldsymbol{R}_{{j}} \partial \boldsymbol{R}_{{i}}}\right| \Psi(\boldsymbol{R})\right\rangle.
\end{equation}
It ignores the wave function derivatives with nuclear coordinates. According to the extended correlated sampling approach, the calculation of Hessian matrix can apply the same ground-state ware function and reutilize measure expectation as done in force calculation.

In this study, all calculations are performed using the ChemiQ software developed by our group\cite{wang2021chemiq}. To evaluate the precision, the classical CCSD method is also performed to obtain ground state energy, force and vibration frequency by PySCF software\cite{sun2018pyscf,sun2020recent}. The STO-3G basis and SLSQP optimizer are used for all calculations. The S$\rm_N$2 reaction adopt the HOMO/LUMO active space (2, 2) to generate fewer qubits and lower circuit. The AIMD time step is 0.2 fs and the difference step is 1.0×10$^{-3}$ Å.

\section{Results and Discussion}
\label{sec3}
Compared to molecular energy calculations, chemical reaction simulation play a more fundamental role in  actual application, which refers to the molecular structure, property and reactivity. The H$^+$ + H$_2$ → H$_2$ + H$^+$ reaction and S$\rm_N$2 (Cl$^-$ + CH$_3$Cl → ClCH$_3$ + Cl$^-$) reaction are one of the most basic reactions in computational chemistry, which are often used as a model to test new methods. Though the hydrogen exchange reaction only includes three atoms, it involves the breakage of old bond and the formation of new bond. Moreover, it is significant for astrophysicist to understand the thermodynamical evolution of early universe. Up to now, there are still some research groups investigating the exchange reaction. For instance, Tomas et al. simulated theoretically the H$^+$ + H$_2$  exchange reaction by different approaches\cite{gonzalez2006detailed}. They found that the reaction dynamics is governed by an insertion mechanism. Recently, the rate constants of H$^+$ + H$_2$  reaction was also studied using a statistical quantum method between T = 5 K and 3000 K\cite{gonzalez2021rate}, which is related to the galaxy formation and evolution.

\begin{figure*}[]
	\centering
	\includegraphics[width=13cm,height=8cm]{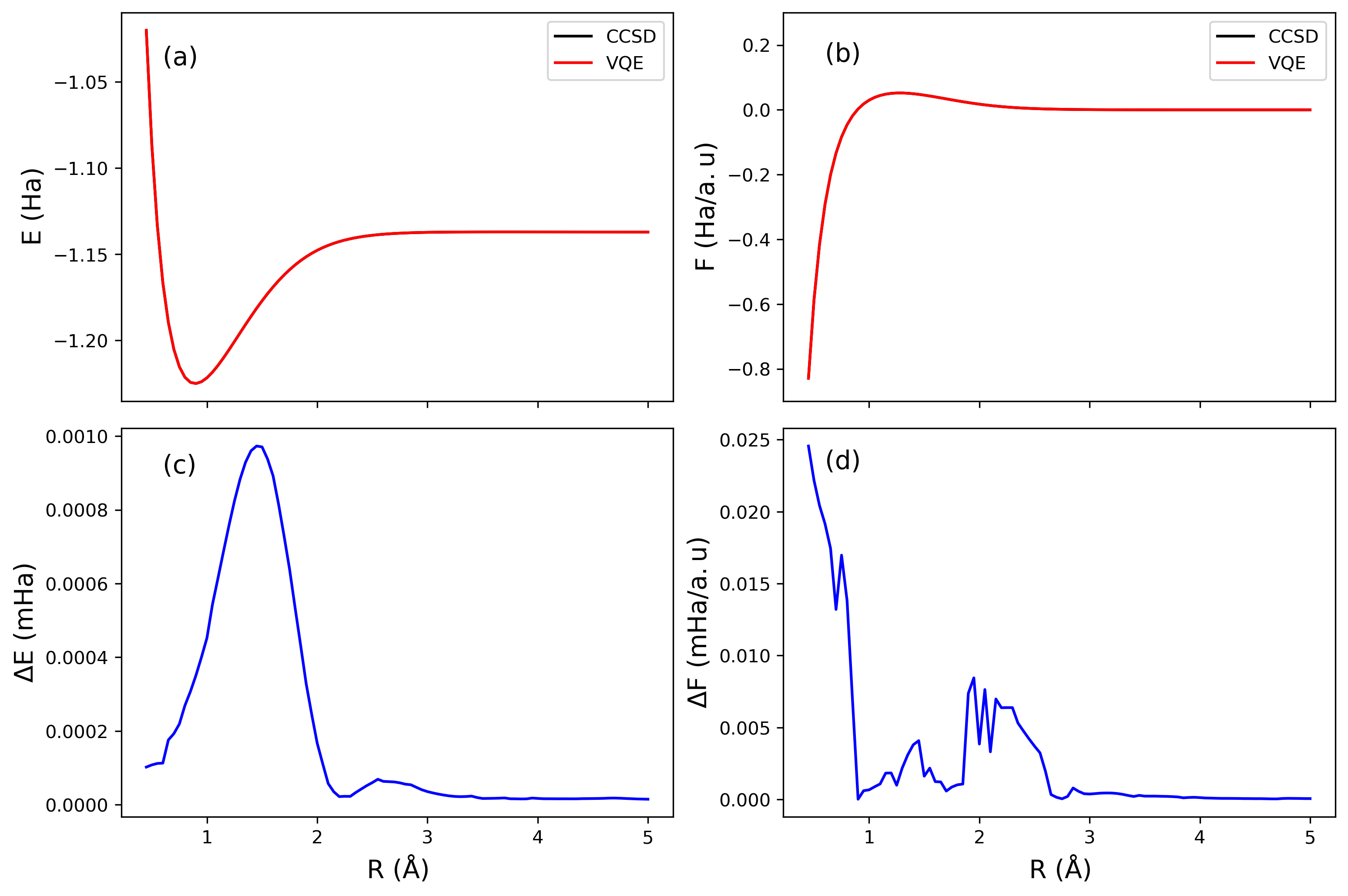}
	\caption{The process of H$^+$ moves close to  H$_2$ molecule from 5.0 Å to 0.45 Å. (a) ground state energies vs distances, (b) forces vs distances, (c) energy deviations $\left|E_{VQE} - E_{CCSD}\right|$ vs distances, (d) force deviations $\left|F_{VQE} - F_{CCSD}\right|$ vs distances.}
	\label{fig1}
\end{figure*}

Before simulating the reaction dynamics of hydrogen exchange reaction, we performed an accurate calculation for H$_3 ^+$ system by CCSD method to evaluate the accuracy of ground-state energies and forces described in Section \ref{sec2}. The ground-state energies obtained by VQE and CCSD for H$^+$ moving close to H$_2$ molecule are shown in Fig. 1(a) and (c). We can see that the  potential energy surface described by VQE algorithm is in agreement with the CCSD method, and maximal deviation occurred at reaction region is about 0.001 mHa far less than the chemical accuracy 1.6 mHa. Fig. 1(b) and (d) displays the forces of H$^+$ and corresponding absolute deviations. The force deviation curve becomes rough and fluctuations probably due to the numerical difference, the maximal error is only about 0.025 mHa/a.u which does not affect the AIMD accuracy. Therefore, these results provide the basis for directly simulating the reaction dynamics of hydrogen exchange reaction.

Since the exchange reaction only includes three particles, it can be described by three relative distances. Fig. 2 shows the colinear reactive collision in the gas phase within 60 fs. The reactive process refers to the H$_A ^+$ ion colliding the hydrogen molecule. It causes a new H$_2$ molecule forms and H$_C ^+$ escapes. In the starting stage, the hydrogen molecule does periodic vibration around the equilibrum bond length, the distance between H$_A ^+$ and H$_B$ decreases uniformly. Next, three particles come into being a complex structure among chemical reaction region. Finally, a new molecule H$_A$H$_B$ forms. Usually, the fairly small time step is necessary for AIMD simulation to reduce integrating error, and total energy conservation can be used to assess the AIMD trajectory quality. Here the maximal total energy drift is observed about 0.92 mHa which happens about 30 fs. Our results demonstrate that the reaction dynamics based on the quantum computing can be simulated accurately.

\begin{figure}[h]
	\centering
	\includegraphics[width=8.5cm,height=7.5cm]{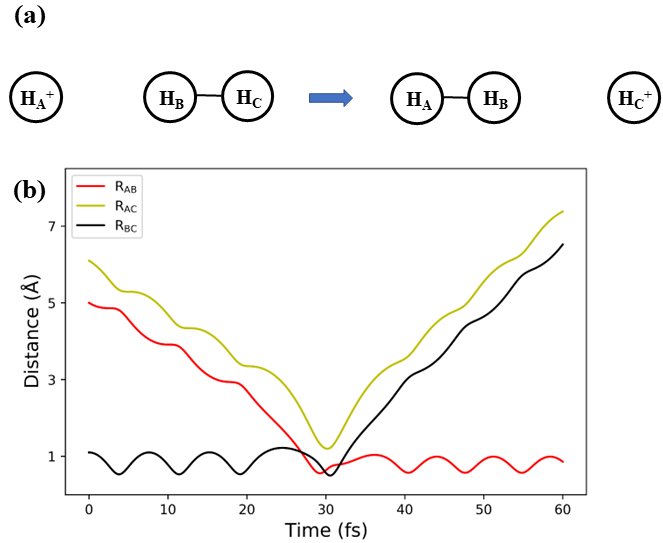}
	\caption{H$^+$ + H$_2$ colinear reactive collision in the gas phase with initial velocity 0.125 Å/fs. (a) Schematic diagram of reaction process, (b) The reaction trajectory obtained by AIMD simulations within 60 fs.}
	\label{fig2}
\end{figure}

\begin{table*}[] 
	\centering  
	\begin{threeparttable}
		\caption{The optimized structures and corresponded frequencies}  
		\begin{tabular} {p{1.7cm}<{\centering} p{1.7cm}<{\centering} p{1.7cm}<{\centering} p{1.7cm}<{\centering} p{2.1cm}<{\centering} p{2.1cm}<{\centering} p{2.1cm}<{\centering}}
			\toprule 
			system   				&${l_0}$(Å)$^a$ & $l$(Å)$^{b1}$ & $l$(Å)$^{b2}$ & Freq(cm$^{-1}$)$^{c1}$ &Freq(cm$^{-1}$)$^{c2}$ &Freq(cm$^{-1}$)$^{c3}$ \\
			\midrule 
			H$_2$   			 		& 1.0      & 0.735 & 0.735   & 5001.9   &  5000.2    &  5201.3  \\
			\midrule
			LiH      			 		& 1.15     & 1.547 & 1.548   & 1680.7   &  1683.3    &  1730.5  \\
			\midrule
			\multirow{3}{*}{H$_3^+$}  	& 1.208    & 0.986 & 0.986   & 3445.9   &  3447.3    &  3526.2  \\
			\multirow{3}{*}{}           & 1.603    & 0.986 & 0.986   & 2116.3   &  2122.3    &  2166.7  \\
			\multirow{3}{*}{}           & 2.566    & 0.986 & 0.986   & 2116.3   &  2115.9    &  2159.2   \\
			\bottomrule 
		\end{tabular}
		$^a$The initial bond length.
		$^{b1}$The optimized bond length by CCSD.
		$^{b2}$The optimized bond length by VQE.
		$^{c1}$The frequency calculated by CCSD.
		$^{c2}$The frequency calculated by VQE.
		$^{c3}$The frequency calculated by VQE with extended correlated sampling approach.
	\end{threeparttable}
\end{table*}


\begin{figure*}[]
	\centering
	\includegraphics[width=12cm,height=7.5cm]{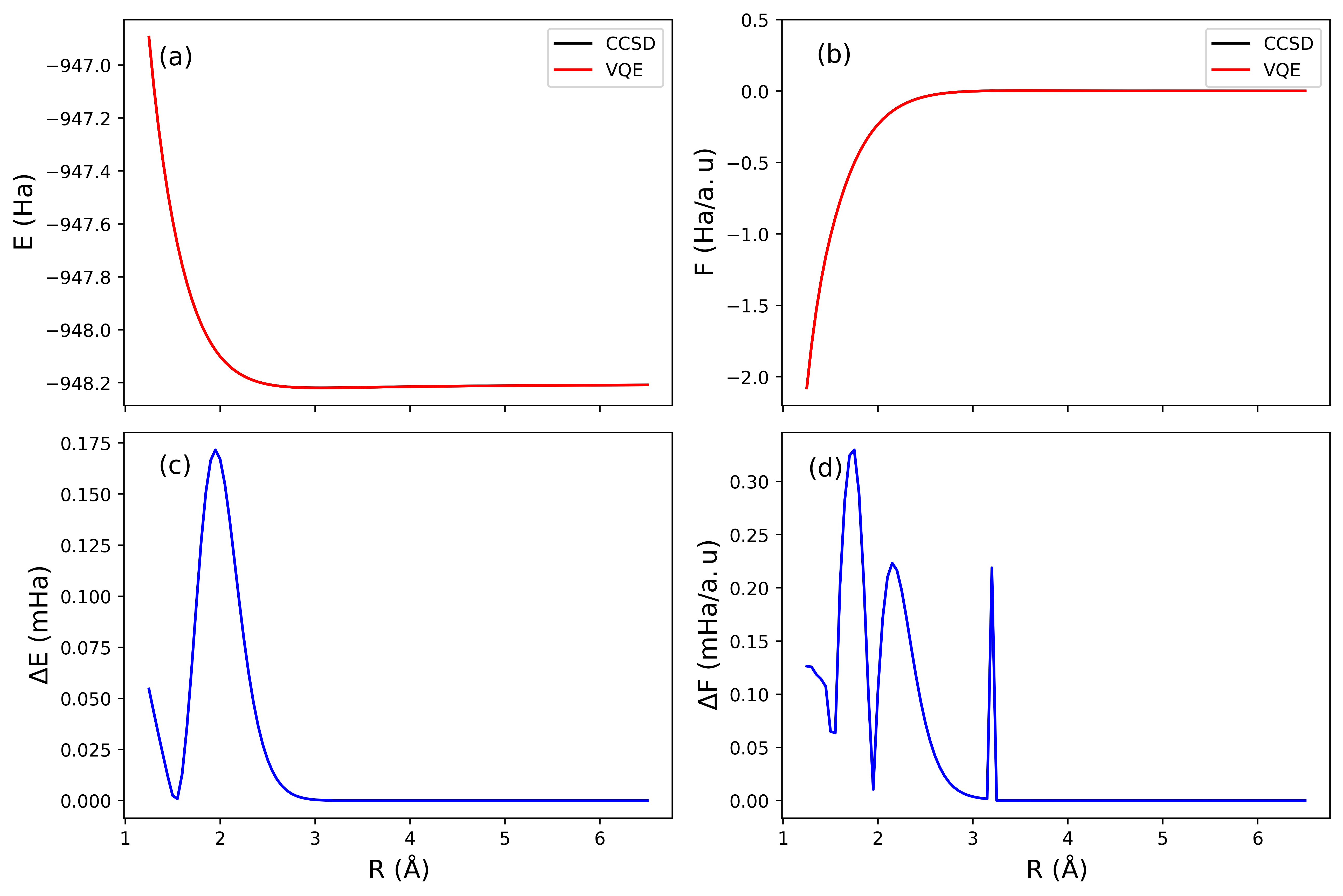}
	\caption{The process of Cl$^-$ closed to CH$_3$Cl molecule from 6.5 Å to 1.25 Å. (a) ground-state energies vs distances, (b) forces vs distances, (c) energy deviations $\left|E_{VQE} - E_{CCSD}\right|$ vs distances, (d) force deviations $\left|F_{VQE} - F_{CCSD}\right|$ vs distances.}
	\label{fig3}
\end{figure*}

Vibration frequency is another important property, which can be applied to calculate the vibration entropy, which determines the transition structure. Usually, the local extreme points of potential energy surface can be approximated by harmonic potential function, and the atomic coupled motions can be separated into individual motions by normal mode analysis. The vibration spectrum involves the calculation of force constant matrix (mass-weighted Hessian matrix), which is complicated and difficult in traditional chemistry. Here, we numerically evaluate the vibration frequencies of H$_2$, LiH and H$_3^{+}$ based on VQE algorithm and extended correlated sampling approach. Firstly, the molecules are optimized to the stationary point from random starting structures by gradient descent method. Next, the vibration frequencies are calculated by Hessian matrix or approximated Hessian matrix. Moreover, the CCSD method is applied to obtain optimized structures and exact frequencies.

\begin{figure*}[]
	\centering
	\includegraphics[width=12cm,height=7.5cm]{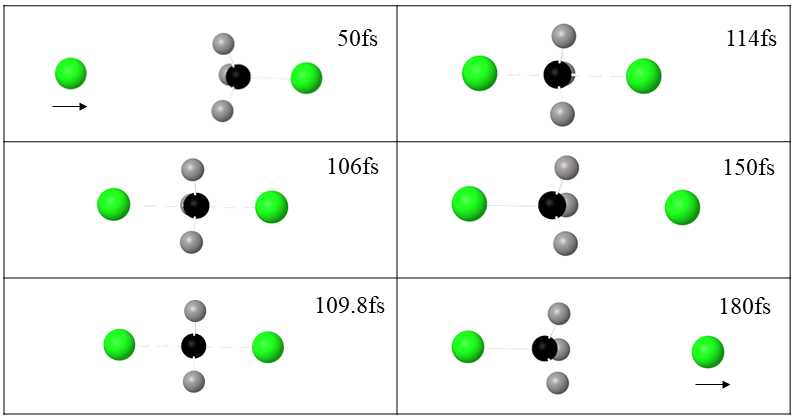}
	\caption{View of the S$_N$2 reaction process from AIMD trajectory.}
	\label{fig4}
\end{figure*}

Table I gives the optimized bond lengths and corresponded vibration frequencies. The equilibrium bond lengths obtained by VQE is consistent with the classical CCSD results, which is the basis for accurately calculating the frequencies.  Since H$_2$ and LiH are both diatomic molecules, there is only one frequency. The frequencies based on  Hessian matrix by VQE algorithm of H$_2$ and LiH molecules are 5000.2 cm$^{-1}$ and 1683.3 cm$^{-1}$ respectively, which is close to the exact results 5001.9 cm$^{-1}$ and 1680.7 cm$^{-1}$. For H$_3^{+}$ molecule, there are three vibration frequencies and the average difference between VQE and CCSD is about 2.6 cm$^{-1}$, it achieves the spectrum precision. Even though the frequency can be highly accurately calculated by this way, the Hessian matrix need to optimize 18N$^2$ ground-state wave functions for N atomic system, which is infeasible with current available hardware. In this perspective, we also investigate the frequency calculation based on approximated Hessian matrix described in section \ref{sec2} "Method" part. It adopts the extended correlated sampling approach by reusing measure probability and ground-state wave function, which can reach 96\% - 98\% relative precision. Therefore, the Hessian matrix based on our correlated sampling approach is more suitable for current quantum hardware.

Transition state (TS) is a first saddle point on potential energy surface and corresponds the maximum energy along the reaction coordinates\cite{schlegel2003exploring,schlegel2011geometry}. Moreover, TS is the basis for calculating the activation energy and determining the reaction path. Up to now, some endeavors have been done to expand the VQE algorithm or its variants to determine the reaction path\cite{kanno2020quantum,sarkar2022modular,azad2022quantum,lim2022quantum,li2022toward}. But almost all studies perform the potential energy surface scanning along the predefined paths obtained from classical chemistry software. It is obvious that the transition structure is unknown in advance universally. Therefore, how to search the TS and determine the reaction path based on quantum algorithm is still an open problem.

To find the transition state of H$^+$ + H$_2$ → H$_2$ + H$^+$ reaction, the starting structure should be close to the transition structure or within quadratic region. Here, we uniformly fetch five points from AIMD trajectory (25 fs - 33 fs) to generate the starting structure. We observe that there is only one imaginary frequency -4316 cm$^{-1}$ and -4724 cm$^{-1}$ at 29 fs and 31 fs respectively. Consequently, the corresponded structures can be adopt as the initial structure to search the transition state. Though searching from different structures, they converge to the same point with wit imaginary frequency -974 cm$^{-1}$ and the bond length R$_{AB}$, R$_{BC}$, R$_{AC}$ are 0.876 Å, 0.876 Å, 1.752 Å respectively. To verify our results, the energy gradient and frequency of the H$_3^+$ transition state are also computed by CCSD method. The CCSD data illustrates that the maximal gradient is only 0.000014 Ha/a.u with one imaginary frequency -993 cm$^{-1}$. It suggests that our approach successfully find the transition state of H$^+$ + H$_2$ reaction.

 The S$\rm_N$2 reaction is a fundamental and significant chemical reaction, has been widely applied in the field of drug discovery and organic synthesis\cite{mikosch2008imaging}. Thus, studying the reaction kinetics in the atomic level is very important for understanding the S$\rm_N$2 microscopic processes. Fig. 3 gives potential energy surface and forces when Cl$^-$ is close to CH$_3$Cl from 6.5 Å to 1.25 Å. We observe that the potential energy surface and forces are consistent with the CCSD data with very low errors 0.17 mHa and 0.33 mHa/a.u respectively. Similarly, the errors will increase when Cl$^-$ is around the CH$_3$Cl at chemical reaction region. In summary, it is safe to directly simulate the process of nucleophile chloride ion (Cl$^-$) attacking electrophile chloromethane (CH$_3$Cl) by AIMD simulation.

Fig. 4 displays a typical reaction process for the S$_N$2 reaction with a initial velocity 0.04 Å/fs for Cl$^-$ along the x axis. We can observe the famous Walden inversion mechanism from the AIMD trajectory. The nucleophilic reagent Cl$^-$ attacks the center carbon atom from the back, then the conformation between carbon atom and three hydrogen atoms changes from umbrella to plane at 109.8 fs, the bond between nucleophilic reagent and carbon atom begin to form and leaving group escapes in the end. To generate the initial guess structure of searching TS, we first calculate the frequencies at 109.8 fs , and then uniformly change the distances between Cl and C by fixing CH$_3$. Finally, the transition state is found with an imaginary frequency -179 cm$^{-1}$. Then classical CCSD method is used to verify the transition structure. The results show that there is only one imaginary frequency -301 cm$^{-1}$ and maximal energy gradient is about 0.0017 Ha/a.u. It suggests that our approach find the approximated transition state of S$\rm_N$2 reaction.

\section{Conclusion}
\label{sec4}
Quantum computer is regarded as a promising tool to solve the electronic structure problem. However with the limited capabilities of the NISQ devices, directly simulating the reaction progress remains a challenge. Here we extend the correlated sampling approach based on VQE algorithm to simulate the reaction dynamics by AIMD simulation.

 Before simulating the chemical reactions H$^+$ + H$_2$ → H$_2$ + H$^+$ and Cl$^-$ + CH$_3$Cl → ClCH$_3$ + Cl$^-$, we first calculate the ground-state energies and forces. The results demonstrate that our approach can accurately describe the potential energy surface and forces in chemical reaction region. Next, we successfully simulate the hydrogen exchange reaction and S$\rm_N$2 reaction based on extended correlated sampling approach. Besides, we discuss geometry optimization, frequency calculation and searching TS by comparing different approaches. The test results suggest that our approach can accurately evaluate the molecular structure, property and reactivity, which is another important expansion of the application of quantum algorithm.


\end{document}